\documentstyle[12pt]{article}
\input epsf
\newlength{\dinwidth}
\newlength{\dinmargin}
\begin{document}

\title{\bf{Hard diffractive electroproduction of heavy flavor vector mesons 
in QCD}}

\author{\bf{Leonid Frankfurt}\thanks{\noindent On leave of
    absence from the St.Petersburg Nuclear Physics Institute, Russia.
    Supported by BSF Grant No. 9200126 }\\ 
{\small \sl School of Physics and Astronomy}\\ {\small \sl Raymond and 
Beverly Sackler Faculty of Exact Sciences}\\
  {\small \sl Tel Aviv University}
\\[0.8cm]
\bf{Aharon Levy}  \\{\small \sl  DESY, Hamburg}\\ {\small \sl and} 
\\{\small \sl School of Physics 
and  Astronomy}\\{\small \sl  Raymond and Beverly Sackler Faculty of Exact 
Sciences}\\
  {\small \sl Tel Aviv University}
}

\date{ }

\maketitle

\begin{abstract}
We outline $QCD$ predictions for the diffractive electroproduction of 
heavy flavor vector mesons and compare them with available data.
\end{abstract}

\setcounter{page}{0}
\thispagestyle{empty}
  
\newpage

\section{ Diffractive electroproduction of heavy flavor vector mesons 
in QCD}

It has been understood recently that at sufficiently large $Q^2$
diffractive electroproduction of vector mesons is calculable in
$QCD$~\cite{Brod94}. The necessary conditions are: i) $1\over 2m_N x$
should be much larger than the radius of the target so that the photon
transforms predominantly into a quark component well before the target;
ii) high $Q^2$ is a necessary but not sufficient condition to guarantee
small interquark distances in the quark component of the wave function of
the projectile photon, and thus enables the applicability of the
assumption of asymptotic freedom;  iii) the initial photon should be
longitudinally polarized to ensure the smallness of the produced quark
configuration when it arrives at the target.  Another potentially
fruitful handle to guarantee the smallness of the interquark distances in
the wave function of the photon is the large mass of the heavy quark
($m_q\to\infty$). M.Ryskin~\cite{Ryskin} suggested that the large mass of
the $c$ quark is a sufficient condition for the applicability of $pQCD$
for the production of $c$ quarks in the with a proton target and to
ensure the applicability of the charmonium model to describe the
light--cone wave function of the produced $J/\psi$ meson. 
 
When the momentum transferred to the target- $t$ tends to zero, the cross
section for the hard diffractive electroproduction of $J/\Psi$ through the
interaction of a longitudinally polarized photon with a target proton at
sufficiently large $Q^2$ has the form in $QCD$~\cite{Brod94}: 
\begin{eqnarray}
\left. {d\sigma^{\gamma^*p\rightarrow Vp}_L\over dt}\right|_{t=0} =
{12\pi^3\Gamma_{V \rightarrow e^{+}e^-} m_{V}\alpha_s^2(Q_1) Q^2 
\eta_V^2 F^2(Q^2) \over \alpha_{EM} (Q^2 + 4m_c^2)^4 N_c^2}  
\Bigg|&& \left(1 + i{\pi\over 2} {d \over d\ln x}\right) 
xG(x,Q_1^2)~\Bigg|^2 \ .
\label{eq1}
\end{eqnarray}
Here, $-Q^2$ is the $"mass"^2$ of the virtual photon , $m_c$ is the
current mass of the produced heavy quark $c$ and $\Gamma_{V \rightarrow
e^{+}e^-}$ is the decay width of the vector meson into a $e^+e^-$ pair. 
$\eta_V={\int {dz\over 2 z(1-z)} \psi_V(z) \over \int dz \psi_V(z)}$ and
$\psi_V(z)$ is the minimal Fock component of the light--cone wave function
of the vector meson with mass $m_V$. $F(Q^2)$ is a $Q^2$ dependent
function to be discussed below. We want to stress that cross sections of
hard diffractive processes as well as any hard processes can be simply
expressed through the distribution of bare quarks and gluons but not
through the distribution of constituent quarks. The factor $x G(x,Q_1^2)$
is the gluon distribution in the proton target.  Within the leading $\ln$
approximations there is ambiguity in the dependence of parton
distributions on the virtuality $Q^2$. To draw attention to this important
point we introduce the variable $Q_1^2$. M. Ryskin suggested~\cite{Ryskin} 
to use $Q_1^2={Q^2+4m_c^2\over 2}$.  Another intuitive
suggestion is to compare parton distributions for the different processes
at the same transverse distances between quarks in the wave function of
the photon~\cite{FKS}.  For moderate $Q^2$ the second suggestion predicts
a somewhat less steep increase of the cross section with decreasing $x$.
In principle due to the difference between the photon "mass" and the mass
of the vector meson, nondiagonal gluon distributions should enter into the
formulae~\cite{Halina}. But at sufficiently small $x$ it should coincide
with the diagonal gluon distributions.  This has been proven by the direct
calculations within the leading $\alpha_s \ln x \ln Q^2$ approximation 
in~\cite{Brod94}. A more general proof will be given elsewhere. By
definition, the factor $F(Q^2)$ includes nonasymptotic effects related to
the photon and the vector meson wave functions. 

When $t$ is different from 0, the vertex for the transition $\gamma^*
\rightarrow V$ should depend rather weakly on the momentum transferred to
the target and should be almost universal for all hard diffractive
processes~\cite{Halina,FKS}. If we denote by $B$ the slope of the
differential cross section $d\sigma/dt$, we may estimate its value at
large $Q^2$ as $B\approx {R^2\over 3}$. Here R is the radius of the gluon
distribution in the target proton. Taking R=0.6fm from the realistic
quark-gluon models of a nucleon we obtain $B=$ 3GeV$^{-2}$. At $Q^2=0$ the
photon-$J/\psi$ vertex leads to an additional contribution to $B$ of
$\approx 0.7$ GeV$^2$~\cite{FKS}. Thus a natural estimate for the slope
of hard diffractive photoproduction of $J/\psi$ is $B(Q^2=0)=4$
GeV$^{-2}$. With increasing $Q^2$ the slope B should decrease to the value
around $3$ GeV$^2$.

\section{Comparison between different approximations}

The cross section of diffractive photo - and electroproduction of $J/\Psi$
mesons has been evaluated by Ryskin~\cite{Ryskin} in terms of the BFKL
approximation and nonrelativistic constituent quark charmonium model for
the $J/\Psi$ meson. Brodsky et al.~\cite{Brod94} used the leading
$\alpha_s \ln {Q^2\over \Lambda_{QCD}^2} \ln {1\over x}$ approximation to
evaluate the cross section of the electroproduction of longitudinally
polarized vector mesons in terms of the minimal Fock component of the
light-cone wave function of vector mesons. In~\cite{FKS} this process has
been evaluated within the leading $\alpha_s \ln {Q^2\over
\Lambda_{QCD}^2}$ approximation of $QCD$ and effects of the quark Fermi
motion were taken into account. In this case conventional leading order
gluon and quark distributions enter, and not asymptotical gluon
distributions as in~\cite{Ryskin} and/or in~\cite{Brod94}. 

Within the approximation which assumes that the minimal Fock component of
the light-cone wave function of the $J/\psi$ can be approximated by the
wave function of a nonrelativistic charmonium model~\cite{Ryskin} the
production of transversely polarized $J/\psi$ can be recalculated through
the cross section of longitudinally polarized $J/\psi$.  If the real part
of the amplitude as well as the quark Fermi motion effects are neglected
the calculation of~\cite{Brod94} leads at small $Q^2$ to the formulae 
of~\cite{Ryskin} 
\footnote{In Ref.~\cite{Brod94}, a factor 4 was missed in
the numerator of Eq.~(\ref{eq1}). We are indebted to Z. Chan for pointing
this out.}.  At the same time at large $Q^2$ the wave functions of all
mesons should approach the universal wave function $z(1-z)$~\cite{CZ} and
the charmonium model approximation would be in variance with $QCD$. (Here
$z$ is the fraction of the momentum of the vector meson carried by the
quark). 
    
The effects of the quark Fermi motion within this model are given by the 
factor $F^2(Q^2)$,
\begin{equation}
F(Q^2)={\int d^3k \left({1\over m_c^2+k^2}\right)^{1\over 4} 
\psi_V(k) {-\Delta\over 4}{1\over \left({Q^2\over 4} + (m_c^2+k^2)\right)} 
\over \int d^3k~ 
{\psi_V(k)  \left({1\over m_c^2+k^2}\right)^{1\over 4} 
\over \left({Q^2\over 4} + m_c^2\right)^2} }\ ,
\label{eq2}
\end{equation}
The substitution $z=0.5(1+{k_3\over \sqrt{k^2+m_c^2}})$ is necessary to
transform diagrams of light cone perturbation theory for the wave function
of $J/\psi$ into nonrelativistic diagrams for the wave function of
$J/\psi$ in charmonium models. The factor $(k^2 + m_c^2)^{-1/4}$ is
included in the definition of the nonrelativistic charmonium wave function
to keep the phase volume to be the same as within the nonrelativistic
approach. Here $\Delta$ is the Laplace operator in transverse momentum
space which acts on the photon wave function. 

In~\cite{FKS} realistic charmonium wave functions calculated from a
power-law~\cite{charm1} and a logarithmic potential~\cite{charm2} were
used. Both functions describe $\Gamma_{J/\Psi \rightarrow e^+e^-}$
reasonably well.  The evaluation leads to the factor $F^2(0)\approx 1/9$. 
The significant suppression of the cross section at moderate $Q^2$ is
caused by the fact that for the realistic charmonium model wave functions
the integral over the quark momenta is slowly convergent at large quark
momenta . 

To visualize the estimate of $F(Q^2=0)$ we decompose it into powers of
${<k^2>\over m_c^2}$:  $F^2(0)=1 -c<{k^2\over m_c^2}>$. Here by
definition $<k^2>={\int k^2 \psi_V(k) d^3 k \over \int \psi_V(k) d^3 k}$.
(Note that for the Coulomb potential- which follows from QCD for the
sufficiently large mass of heavy flavor $\int k^2 \psi_V(k) d^3 k
=\infty$.  So this decomposition is inapplicable for the production of
sufficiently heavy flavors where applicability of $pQCD$ is most
reliable.) The coefficient $c$ has been evaluated in~\cite{RRLM} as $c=4$
and in~\cite{FKS} as $c=6{2\over 3}$ . This difference is due to the
leading and next to leading powers of $k^2\over m_c^2$ in the evaluation
of the operator $\Delta$ both of which are taken into account
in~\cite{FKS} but not in~\cite{RRLM}. For the numerical estimates Ryskin
et al.~\cite{RRLM} used a gaussian type approximation to the wave
function of the $J/\psi$ and suggested $<k^2/m_c^2>\approx {1/8}$. 
Within this approximation $F^2(0)\approx 0.5$ for $c=4$~\cite{RRLM} and
$F^2(0)\approx {1\over 6}$ for $c=6{2\over 3}$~\cite {FKS} . The
numerical approximations leading to the difference between the
calculations of~\cite {FKS} and of~\cite{RRLM} are discussed in detail 
in~\cite{FKS}. This problem deserves further investigation. 

If the dependence of the interquark potential on the mass of the quark (on
the running coupling constant, on the running quark mass etc ) can be
neglected, the cross section would be $\sigma \approx {1\over m_c^6}$.
Within the range of constituent quark masses between $m_c=1.48-1.54$ GeV
which are explored in the realistic charmonium models, and at the same
time not far from the bare mass of the $c$ quark the cross section changes
within $42\%$. Calculations within the realistic charmonium models produce
weaker dependence on $m_c$~\cite{FKS}.

\section{Restoration of flavor symmetry }

For the production of vector mesons with $M_V^2 \ll Q^2$ by longitudinally
polarized photons, all dependence on the quark masses and thus on flavor
is contained in the light-cone wave functions of the vector mesons only.
Color exchange is flavor blind and it is legitimate to neglect the masses
of current quarks in the hard scattering amplitude .  $SU(4)$ predicts
that at sufficiently large $Q^2$ the ratio of the production cross section
for charmed mesons to that of the $\rho^0$ is $J/\Psi:\rho^0 =8 : 9$ . The
production of $J/\Psi$ should be additionally enhanced at large $Q^2$ by
the larger probability for a smaller object to annihilate into small
object.  Experimentally, this ratio is suppressed at small $Q^2$ as
compared to the prediction of $SU(4)$ symmetry by a factor of $\approx
25$.  Such a suppression is naturally explained within the above discussed
model and it should slowly disappear with increasing $Q^2$~\cite{FKS}.
Thus QCD predicts \footnote{Note however that this view is not shared by
M.Ryskin.} a dramatic increase of the $J/\Psi/\rho^o$ ratios at large
$Q^2$~\cite{Halina}. 

At very large $Q^2\gg M_V^2$, the $q\bar q$ wave functions of all mesons
should converge to a universal asymptotic wave function~\cite{CZ}. In this
limit, further enhancement of heavy flavor production is expected leading
to $J/\Psi : \rho^0 = (8*3.4) : 9$~\cite{Halina}. The analysis of the
factor $F(Q^2)$ discussed above shows that the observation of the full
restoration of the predictions of $SU(4)$ symmetry for the $J/\Psi$--meson
production would require extremly high values of $Q^2$.

\section{Comparison with the HERA data}

The presently published results of the HERA experiments on diffractively 
produced $J/\Psi$ is for $Q^2$ = 0~\cite{H1,ZEUS}. Some preliminary 
results in the electroproduction region ($<Q^2> = 18$ GeV$^2$) have been 
presented by H1 at the EPS meeting in Brussels~\cite{H1br}.
\vspace{-3cm}
\begin{figure}[h]
\epsfxsize12.0cm
\centerline{\epsffile{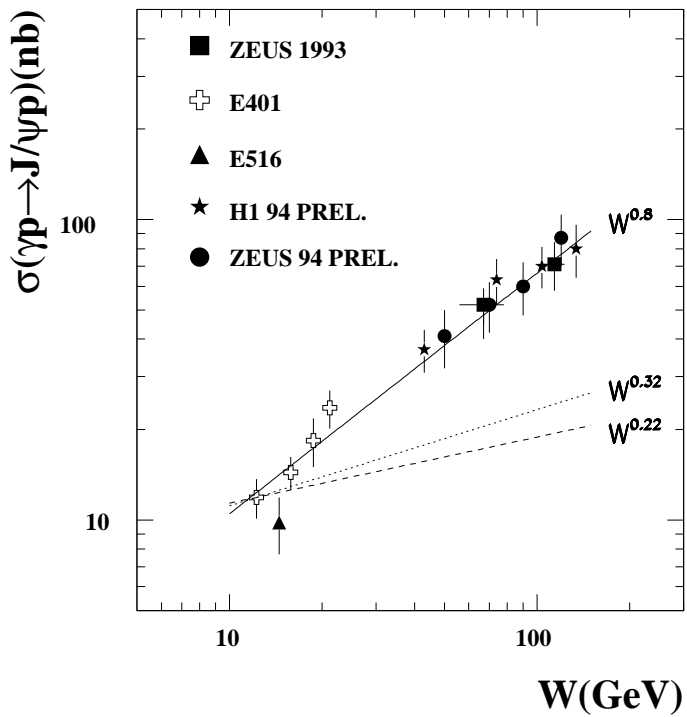}}
\vspace{-3.3cm}
\caption
{A compilation of $J/\Psi$~elastic cross sections in photoproduction.  
The solid line is a $W^{0.8}$ behaviour while the dotted and dashed lines 
are the predictions following the Donnachie
and Landshoff model{[12]} without ($W^{0.32}$) and with ($W^{0.22}$)
shrinkage.}
\end{figure}

A compilation of the photoproduction cross section for the reaction
$\gamma p \to J/\Psi p$ as function of the $\gamma p$ center of mass
energy $W$ is presented in figure 1. One can see a fast increase of the
cross section with $W$, approximately like $W^{0.8-0.9}$. Note that this
increase is also seen in the HERA data alone, thus independent of
possible relative normalizations between the low and high energy
experiments. Such a fast increase of the cross section with $W$ is
unexpected within the non--perturbative two--gluon exchange of Donnachie
and Landshoff~\cite{DL1}, which predict a rise like $W^{0.22}$ after
taking shrinkage into account.  The rise is consistent with a model
calculated within leading $\ln x$ approximation which connects the rise
to the behaviour of the gluon density with decreasing $x$~\cite{Ryskin}. 

The preliminary data from the electroproduction of $J/\Psi$~\cite{H1br}
indicate that the $Q^2$ dependence is steeper than that expected from
GVDM. However there are two QCD effects, which tend to slow down the
dependence on $Q^2$. One effect is that the gluon density increases with
$Q^2$ at small $x$. Numerically, the factor $\alpha_s^2(Q^2) x^2
G_N^2(x,Q^2)$ in Eq.~(\ref{eq1}) is $\propto Q^n$ with $n \sim 1$. In
addition a {\it preasymptotic} effect comes from the factor $F(Q^2)$ of
Eq. (\ref{eq2}). The combination of these two effects results in a
relatively slow decrease with $Q^2$ as compared to ${1\over
(Q^2+4m_c^2)^3}$. This could be checked once higher statistics data are
available.

\section{Summary}

To summarize, the investigation of exclusive diffractive processes appears
to be an effective method to measure the minimal Fock state $q\bar q$
component of the light-cone wave functions of vector mesons as well as the
light-cone wave functions of any small mass hadron system having angular
momentum one. 

\section*{Acknowledgements}
We are indebted to M.Ryskin for reading the text and for his valuable
comments.


\end{document}